\documentclass[amsmath,amssymb,aps,prb,11pt]{revtex4-2}
\usepackage[dvips]{graphicx}
\usepackage{subfigure}
\usepackage{dcolumn}
\usepackage{bm}
\usepackage{hyperref}
\usepackage{array}
\usepackage{upgreek}
\usepackage[usenames]{color}
\usepackage[utf8]{inputenc}
\usepackage[T1]{fontenc}
\usepackage{mathptmx}

\newcolumntype{P}[1]{>{\centering\arraybackslash}p{#1}}
\begin{document}
\preprint{AIP/123-QED}
\title{Emergent impedance due to antiferromagnetic domain wall dynamics}
\author{Yuta Yamane,$^{1,2}$ Jotaro J. Nakane,$^3$ Yasufumi Araki,$^4$ and Jun'ichi Ieda$^{4}$}

\affiliation{$^1$Frontier Research Institute for Interdisciplinary Sciences, Tohoku University, Sendai 980-8578, Japan}
\affiliation{$^2$Research Institute of Electrical Communication, Tohoku University, Sendai 980-8577, Japan}
\affiliation{$^3$Department of Physics, Nagoya University, Nagoya 464-8602, Japan}
\affiliation{$^4$Advanced Science Research Center, Japan Atomic Energy Agency, Tokai 319-1195, Japan}
\date{\today}

\begin{abstract}
We theoretically investigate emergent impedance induced by domain-wall dynamics in antiferromagnets.
Emergent impedance, arising from a combined action of spin-transfer torque and spinmotive force, was previously predicted and observed in spiral magnets.
Here we develop a formalism for the electrical response of an antiferromagnetic domain wall under ac currents, and obtain analytical expressions for the resulting emergent impedance.
We find that two dynamical modes play separate roles in the emergent impedance:
Translational motion of the domain-wall center generates a contribution proportional to its velocity, analogous to that arising from the corresponding motion of a spiral magnet.
Another contribution, unique to antiferromagnetic domain walls, originates from the time-dependent canting of the sublattice magnetizations localized within the moving domain wall, whose magnitude is inversely proportional to the antiferromagnetic exchange coupling constant.
The competition between these two distinct contributions determines the sign and magnitude of the imaginary part of the emergent impedance at sub-resonant frequencies.
Our results provide a fundamental insight into electron transport in antiferromagnets, and open avenues for novel antiferromagnet-based spintronics devices.
\end{abstract}

\maketitle
\section{Introduction}
Since the prediction of staggered magnetic order \cite{neel} and its subsequent experimental confirmations \cite{shull}, antiferromagnets (AFMs) have occupied an important place in a wide range of fundamental physics.
Their lack of macroscopic magnetization, however, makes it difficult to access them by conventional magnetic means effective for ferromagnets.
This limitation has long hindered the technological exploitation of AFMs.
In recent years, the emerging field of antiferromagnetic spintronics \cite{review1,review2,review3}, which seeks to harness the distinctive physical properties of AFMs in spintronics devices, has begun to reshape the conventional view of their technological potential.

It has been shown that magnetic order parameters in AFMs can be manipulated by electric current via spin-transfer torque (STT)\cite{nunez,uraz,xu,helen2010,hals,swaving,linder,tveten,cheng2014,yamane2016-stt,barker,nakane} and spin-orbit torques \cite{jakob,wadley,helen2016,moriyama,shiino,chen,tsai,takeuchi2021,higo,zhang,pal,takeuchi2025}.
The reciprocal effects of these current-induced torques, in turn, enable dynamical order parameters to generate so-called emergent electromagnetic fields acting on conduction electrons \cite{cheng2012,helen2015,okabayashi,yamane2016-smf}.
A direct manifestation of the emergent electric field is an induced electric voltage, or spinmotive force (SMF) \cite{berger,volovik,stern,barnes,yang,hals2015}.
These phenomena are well established in ferromagnets and have been observed in a variety of ferromagnetic textures.  
AFMs can similarly host topologically nontrivial textures of their order parameters, such as domain walls (DWs) \cite{bary,papa,bode} and skyrmions \cite{barker,bogd}.
The interplay between such AFM textures and electric currents/voltages is key to AFM-based technologies.

Recently, an electrical impedance emerging from DW dynamics in an AFM has been reported \cite{yokouchi-arxiv}.
This impedance, or emergent impedance, arises from a sequential action of the current-induced torques and the SMF:
when the magnetic order parameter is driven by a time-varying electric current, its dynamics generate a SMF in addition to the applied electric voltage.
This whole process modulates the electrical impedance of the magnet.
This phenomenon was originally predicted \cite{nagaosa} and later observed \cite{yokouchi,kitaori1,kitaori2,kitaori3,matsushima} in spiral magnets and ferromagnets, particularly focusing on inductive responses.
Theoretical studies have discussed the contributions to the emergent impedance from different excitation modes of spiral textures \cite{kurebayashi,anan}, electron correlations \cite{oh}, Joule heating effects \cite{furuta1,furuta2} and spin-orbit couplings \cite{ieda,yamane2022,araki}.
The discovery of the emergent impedance has opened possibilities for revolutionizing electronic circuit elements, such as inductors, through the quantum-mechanical spintronics effects.
The extension of the concept of emergent impedance to AFMs \cite{yokouchi-arxiv} not only highlights a new aspect of fundamental electron transport in AFMs, but also offers greater varieties in functionalities and material platforms for emergent-impedance devices.
Accordingly, a dedicated theoretical framework is now required to elucidate the mechanisms governing emergent impedance in AFMs and to guide its further experimental explorations.

In this work, we present a theoretical formalism of emergent impedance induced by DW dynamics in AFMs.
In Ref.~\cite{yokouchi-arxiv}, we provided an analysis of the experiment by employing a model that is appropriate and sufficient to explain the specific experimental observations:
namely, the strong AFM exchange-coupling limit, where canting of the sublattice magnetizations is excluded, within a low-frequency regime (up to $\sim$kHz in the experiment) far below the characteristic frequencies of the material.
Here, we develop a more general theory applicable to broader frequency ranges and to a wider class of AFM systems, including synthetic AFMs.
We find that two distinct mechanisms contribute to the emergent impedance:
one arising from translational DW motion and the other from the time-dependent canting.
The former contribution is proportional to the translational velocity of the DW center, analogous to the corresponding effect in spiral magnets and ferromagnetic DWs, whereas the latter scales inversely with the AFM exchange coupling constant, an effect unique to AFM DWs.
The competition between these two contributions determines the sign and magnitude of the imaginary part of the emergent impedance in the low-frequency regime.
We examine the frequency characteristics of the emergent impedance up to and beyond its resonance frequency, which allows direct comparison with future experiments.

\section{Model}
We consider an AFM metal composed of two magnetic sublattices, $A$ and $B$, each with the same saturation magnetization $M_{\rm S}$.
The classical unit vector $\vec m_A(\vec r, t)$ denotes the local magnetization direction in sublattice $A$, and $\vec m_B(\vec r, t)$ is similarly defined for sublattice $B$.
Both $\vec m_A(\vec r,t)$ and $\vec m_B(\vec r,t)$ are treated as continuous fields in real space, following the coarse-graining of the underlying atomistic structure \cite{lifshitz}.
Throughout this work, we assume spatial uniformity in the $yz$ plane, i.e., $\vec m_{A,B}(\vec r,t)=\vec m_{A,B}(x,t)$, since our focus will later be on one-dimensional DWs, which we take (without loss of generality) to extend along the $x$ axis.

The magnetic energy density $U$ associated with the sublattice magnetizations is assumed to be given by
\begin{eqnarray}
U &=& J_{\rm AF} \vec m_A \cdot \vec m_B + A_1 \left[ \left( \partial_x \vec m_A \right)^2 + \left( \partial_x \vec m_B \right)^2 \right] \nonumber \\ && - A_2 \left( \partial_x \vec m_A \right) \cdot \left( \partial_x \vec m_B \right) - K (m_{Az}^2 + m_{Bz}^2) ,
\label{u}
\end{eqnarray}
where $ J_{\rm AF}(>0)$ is the homogeneous AFM exchange coupling constant, $A_1$ and $A_2$ are the intra- and intersublattice exchange stiffnesses, respectively, and $K(>0)$ is the uniaxial anisotropy constant.
In the presence of an electric current flowing along the $x$ direction, the dynamics of $\vec m_i$ $(i=A,B)$ are governed by the coupled Landau-Lifshitz-Gilbert (LLG) equations,
\begin{equation}
{\cal D}_t \vec m_i = - \vec m_i \times \left(\vec \omega_i - \alpha {\cal D}_t^\beta \vec m_i \right) ,
\label{llg}
\end{equation}
where $\alpha$ is the Gilbert damping constant.
The effective fields $\vec \omega_i$ are defined as $\vec \omega_i = - \frac{\gamma}{\mu_0 M_S}\frac{\delta U}{\delta \vec m_i}$ with $\gamma$ and $\mu_0$ denoting the gyromagnetic ratio and the vacuum permeability, respectively.
The STT, which accounts for the effect of the electric current, is incorporated as non-dissipative and dissipative torques via \cite{swaving,yamane2016-stt}
\begin{equation}
{\cal D}_t = \partial_t + u \partial_x , \qquad
{\cal D}_t^\beta = \partial_t + \frac{\beta}{\alpha} u \partial_x ,
\end{equation}
respectively.
Here, $u$ is a parameter with dimensions of velocity, proportional to the electric current density $j_{\rm c}$, and $\beta$ is a phenomenological dimensionless constant describing the dissipative part of the STT \cite{tatara,zhang-s,barnes2005,tserkov,garate}.
We introduce another phenomenological dimensionless constant $\mu$ such that
\begin{equation}
u=\frac{\gamma\hbar}{2e\mu_0 M_{\rm S}} \mu j_{\rm c} ,
\label{uj}
\end{equation}
where $e(>0)$ is the elementary charge and $\hbar$ is the reduced Planck constant.
The physical interpretation of $\mu$ will be touched upon later.

We introduce the N\'eel order vector $\vec n = \frac{\vec m_A - \vec m_B}{2}$ and the canting moment $\vec m = \frac{\vec m_A + \vec m_B}{2}$.
Throughout this work, we assume $|\vec n|\simeq 1$ and $|\vec m|\ll 1$, which holds when the effective field due to the exchange coupling $J_{\rm AF}$ is the dominant one in Eq.~(\ref{llg}) in the presence of STT, a condition typically satisfied with most AFM materials.
Under this condition, together with the assumption $\alpha \ll1$, a perturbation expansion of the coupled LLG equations (\ref{llg}) up to first order of $J_{\rm AF}^{-1}$ leads to an expression of $\vec m$ as a function of $\vec n$ \cite{bary,andreev}:
\begin{equation}
\vec m = - \frac{\vec n \times {\cal D}_t \vec n}{\omega_E} ,
\label{m}
\end{equation} 
and to a closed equation of motion for $\vec n$ \cite{hals,swaving,yamane2016-stt,nakane}:
\begin{equation}
\vec n \times \left[ \left( {\cal D}_t^2 - c^2 \nabla^2 \right) \vec n - \omega_E \left( \omega_K n_z \vec e_z - \alpha {\cal D}_t^\beta \vec n \right) \right] = 0 ,
\label{eom}
\end{equation}
where $\omega_E = \frac{2\gamma J_{\rm AF}}{\mu_0 M_S}$, $c^2 = \frac{2\gamma (2A_1+A_2) \omega_E}{\mu_0 M_S}$, and $\omega_K = \frac{2\gamma K}{\mu_0 M_S}$.

The conduction electrons, on the other hand, are assumed to be described by the $4\times4$ one-body Hamiltonian
\begin{equation}
H = \left( \begin{array}{cc} t (\vec p) & t' (\vec p) \\ t'^{*} (\vec p) & t (\vec p) \end{array} \right) + \left( \begin{array}{cc} J \vec \sigma \cdot \vec m_A (x,t) & 0 \\ 0 & J \vec \sigma \cdot \vec m_B (x,t) \end{array} \right) .
\label{h}
\end{equation}
The diagonal and off-diagonal blocks correspond to the intra- and intersublattice parts, respectively, with the upper-left (bottom-right) block referring to sublattice $A$ ($B$).
The first term in Eq.~(\ref{h}) represents the kinetic part, where $t$ and $t'$ denote the intra- and intersublattice electron hopping energies, respectively, and $\vec p$ is the electron momentum operator.
The second term describes the exchange coupling between the conduction-electron spin and the sublattice magnetizations, where $J$ is the coupling constant and $\vec \sigma$ is the vector of Pauli matrices representing the conduction-electron spin operator.

Through the exchange couplings in Eq.~(\ref{h}), the dynamics of the conduction electrons and the sublattice magnetizations are interdependent.
The time-dependent sublattice magnetizations generate the emergent electric field acting on the conduction electrons \cite{cheng2012,helen2015,okabayashi,yamane2016-smf}.
Owing to the symmetry of the AFM under interchange of the two sublattices, i.e., under reversal of $\vec n$, observable effects must be expressed as even functions of $\vec n$.
This symmetry requirement restricts the allowed form of the net emergent electric field $\vec E$ to \cite{yamane2016-smf}
\begin{equation}
\vec E = E \vec x \equiv \frac{\hbar\mu}{2e} \left\{ \beta \partial_t \vec n \cdot \partial_x \vec n + f \left[ {\cal O}(\vec m) \right] \right\} \vec x ,
\label{e}
\end{equation}
up to first order in $\vec m$ and in the spatiotemporal derivatives.
Here, $\vec x$ is the unit vector along the $x$ axis and
\begin{equation}
f \left[ {\cal O} (\vec m) \right] =\vec n \times \partial_t \vec n \cdot \partial_x \vec m + \vec n \times \partial_t \vec m \cdot \partial_x \vec n .
\label{f}
\end{equation}
From the reciprocity between the STT and the emergent electric field, $\mu$ and $\beta$ in Eq.~(\ref{e}) coincide with those appearing in Eq.~(\ref{llg}).
In previous studies, $f\left[ {\cal O} (\vec m) \right]$ has been largely ignored since it vanishes in the strict AFM limit of $\omega_E\rightarrow\infty$.
In this work, however, we will show that $f\left[ {\cal O} (\vec m) \right]$ can produce a  qualitatively and quantitatively important contribution to the emergent impedance.

An analytical expression and/or intuitive physical interpretation for the phenomenological parameter $\mu$ may be available when the conduction-electron Hamiltonian (\ref{h}) is sufficiently simple.
For instance, in the limit of $J\gg |t'(\vec p_F)|$ with $\vec p_F$ the Fermi momentum, which can be a reasonable approximation for some synthetic AFMs, the system effectively reduces to two FMs.
As shown in Ref.~\cite{yamane2016-smf}, $\mu$ in this case coincides with the spin polarization of each FM, namely each sublattice.
In general, however, $\mu$ depends intricately on the electronic band structure and the location of the Fermi surface \cite{nakane}.
Note also that, in Eq.~(\ref{llg}), we neglected the correction to the STT arising from the back action of the emergent electric field, which would lead to an spatially-dependent effective damping term \cite{zhang2009}.
This approximation is justified because the spin current generated by the emergent electric field is typically much smaller than the spin current associated with the externally applied current.

\begin{figure}[t]
\centering
\includegraphics[width=8.5cm, bb=0 0 1752 649]{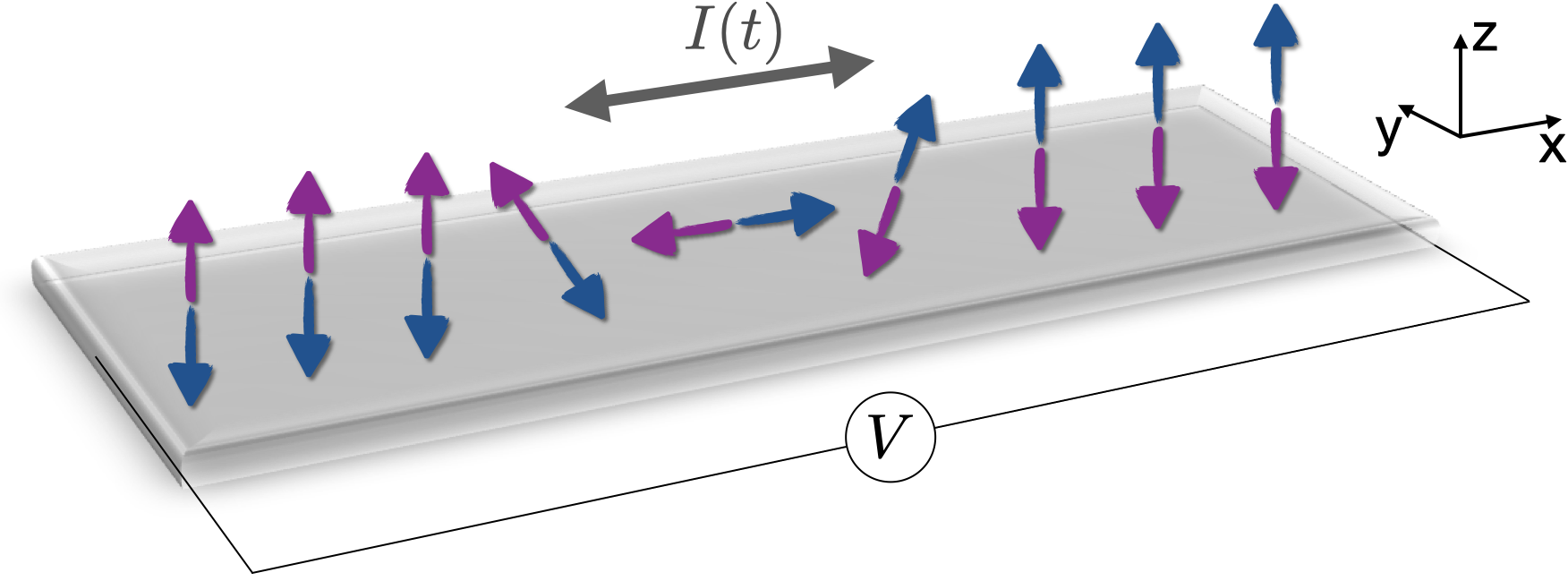}
\caption{
Schematic of our model system, where the DW dynamics driven by the time-varying electric current $I(t)$ generates the SMF $V$.
Both sublattice magnetizations are defined at each spatial point as a result of the coarse graining.
The model can be applied to synthetic as well as intrinsic AFMs.
}
\end{figure}

\section{Domain-wall dynamics and spinmotive force}
In this section, we discuss current-driven DW dynamics and the resulting SMF (Fig.~1).
The N\'eel vector is expressed in polar coordinates as $\vec n\simeq(\sin\theta\cos\phi,\sin\theta\sin\phi,\cos\theta)$, where $|\vec n|\simeq1$ has been used.
We adopt the following dynamical DW ansatz;
\begin{eqnarray}
\theta(x,t) &=& 2\tan^{-1}\left[\exp\left(q\frac{x-X(t)}{\lambda}\right)\right], \label{theta} \\
\phi(x,t) &=& \Psi(t). \label{phi}
\end{eqnarray}
Here, $q = \frac{1}{\pi}\int_{-\infty}^\infty dx \partial_x \theta=\pm1$ is the topological charge of the DW, $\lambda$ is the DW width.
The collective coordinates $X(t)$ and $\Psi(t)$ describe the translational and precessional motions of the DW, respectively, with the former schematically depicted in Fig.~2~(a).
Substituting this ansatz into Eq.~(\ref{eom}) and phenomenologically introducing a pinning potential $U_{\rm pin}$ for $X$, the equations of motion for $(X,\Psi)$ are obtained as
\begin{eqnarray}
\frac{1}{\omega_E}d_t^2 X + \alpha d_t X &=& \beta u + \frac{1}{\omega_E}d_t u -\frac{dU_{\rm pin}}{dX}, \label{eom_x} \\
\frac{1}{\omega_E}d_t^2 \Psi + \alpha d_t \Psi &=& 0 . \label{eom_psi}
\end{eqnarray}
Notice that the dynamics of $X$ and $\Psi$ are decoupled \cite{yamane2016-stt,nakane}.
The STT effects, which depend on $u$, appear only in Eq.~(\ref{eom_x}) and are absent in Eq.~(\ref{eom_psi}).
As a result, $d_t\Psi$ remains identically zero and thus the precessional motion can be ignored.
Consequently, Walker breakdown, a universal phenomenon well-known in ferromagnetic DW motion, in which the translational motion slows down upon the onset of precessional motion \cite{walker}, does not occur in the AFM case.

\begin{figure}[t]
\centering
\includegraphics[width=8.5cm, bb=0 0 1776 965]{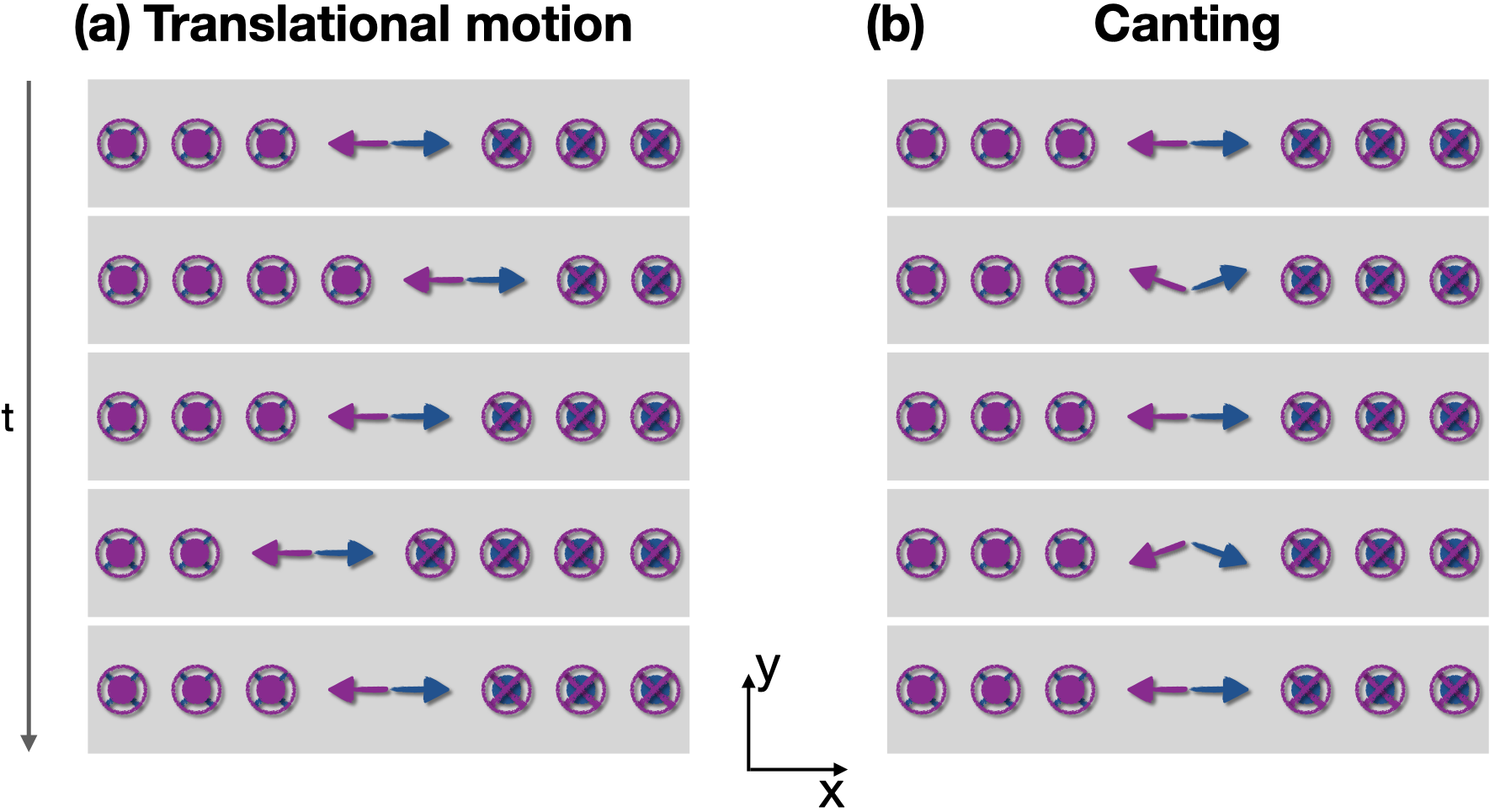}
\caption{
Schematic illustrations of the two dynamical mode of a DW: (a) translational motion and (b) time-dependent canting.
The two modes are coupled [see Eq.~(\ref{m_dw})], and a general DW dynamics consists of a combination of them.
}
\end{figure}

For the pinning potential, we assume a harmonic form
\begin{equation}
U_{\rm pin} = \frac{\omega_{\rm pin}}{2} X^2 ,
\label{u_pin}
\end{equation}
where $\omega_{\rm pin}$ is the characteristic angular frequency of the potential.
Equation~(\ref{u_pin}) provides a good description for a DW trapped near a local minimum of the random potential created by imperfections such as impurities and inhomogeneities.

Equations (\ref{eom_x}) and (\ref{eom_psi}) describe the dynamics of the N\'eel vector $\vec n$ through the collective coordinates.
The moving DW is also accompanied by the canting moment $\vec m$ varying in time [Fig.~2~(b)].
Substituting the DW structure given by Eqs.~(\ref{theta}) and (\ref{phi}) into Eq.~(\ref{m}) yields
\begin{equation}
\vec m = \frac{ q \sin\theta }{ \omega_E \lambda } \left( - d_tX + u \right) \left( \vec x \sin\Psi - \vec y \cos\Psi \right) .
\label{m_dw}
\end{equation}
The magnitude $|\vec m|$ is appreciable only within the DW region, and vanishes far from the DW where $\theta=0$ and $\pi$.

The dynamics of the DW described by Eqs.~(\ref{theta}), (\ref{phi}) and (\ref{m_dw}) give rise to the emergent electric field according to Eq.~(\ref{e}).
Following (the one-dimensional version of) Poisson's equation, this field induces the SMF across the DW by $V=\int_{-\infty}^\infty dx \ E$ \cite{ohe}.
The explicit expression is given by
\begin{equation}
V = \frac{\hbar\mu}{e\lambda} \left[ -\beta d_t X + \frac{1}{\omega_E} \left( -d_t^2 X + d_t u \right) \right] .
\label{smf}
\end{equation}
The first term, proportional to $d_tX$, arises from the translational motion of the DW and was derived previously \cite{yamane2016-smf,yokouchi-arxiv}.
A similar effect is also known for motion of ferromagnetic DWs \cite{duine} and spiral structure \cite{ieda,kurebayashi}.
(In the context of current-driven dynamics, a spiral magnet can be treated as a chain of ferromagnetic DWs \cite{hals2013}.)
In contrast, the second and third terms, proportional to the acceleration of the DW center and to the time derivative of the electric current density, respectively, have not been discussed before.
They originate from the time-dependent canting moment $\vec m$ via the second term in Eq.~(\ref{f}), which is an effect unique to AFMs with no direct analogue in ferromagnets.
In the strict AFM limit $\omega_E\rightarrow\infty$, these canting-induced contributions vanish because $\vec m\rightarrow0$.
Equation~(\ref{smf}) encapsulates one of the central findings of this work:
the SMF generated by a dynamical DW in an AFM consists of the two distinct contributions, one from translational DW motion governed by the dynamics of $\vec n$, and the other from the time-dependent canting moment $\vec m$ localized within the moving DW.

Despite the assumed smallness of $1/\omega_E$, the second and third terms in Eq.~(\ref{smf}) can be comparable to, or even exceed, the first term.
This is because, besides the simple fact that each term exhibits different dynamics under a given current, $\beta$ can be much smaller than unity.
In what follows, we investigate the emergent impedance associated with the SMF in Eq.~(\ref{smf}).

\section{Emergent impedance}
We consider an ac electric current density $j_{\rm c}(t)=j_\omega e^{i\omega t}$, and seek a solution to Eq.~(\ref{eom_x}) of the form $X(t)=X_\omega e^{i\omega t}$.
Here, $j_\omega$ denotes a real amplitude, while $X_\omega$ is in general complex.
Equation~(\ref{eom_x}) is solved for $X_\omega$ as
\begin{equation}
	X_\omega = - \frac{1}{ \frac{\omega^2}{\omega_E} - \omega_{\rm pin} - i \alpha \omega }
	                            \left( \beta + \frac{ i \omega }{ \omega_E } \right) u_\omega .
\label{x_omega}
\end{equation}
Here, $u_\omega$ is the real amplitude defined through $u(t)=u_\omega e^{i\omega t}$, related to $j_\omega$ via Eq.~(\ref{uj}).
With the SMF expressed as $V(t)=V_\omega e^{i\omega t}$, the emergent impedance, defined as $Z_\omega = V_\omega/I_\omega$ with $I_\omega=S j_\omega$ representing the total-current amplitude and $S$ the cross-sectional area, is obtained as
\begin{eqnarray}
Z_\omega
&=& i\omega\left(\frac{\hbar\mu}{e}\right)^2 \frac{\gamma}{2\mu_0 M_{\rm S}\lambda S} \nonumber \\
&& \times \left[ \frac{1}{ \frac{\omega^2}{\omega_E} - \omega_{\rm pin} - i \alpha \omega } \left( \beta + \frac{i\omega}{\omega_E} \right)^2 + \frac{1}{\omega_E} \right] .
\label{z}
\end{eqnarray}
Equation~(\ref{z}) has a Lorentz-like form, with the resonance frequency approximately given by $\sim\sqrt{\omega_E\omega_{\rm pin}}$ under the assumption $\alpha \ll 1$.
In the strict AFM limit $\omega_E\rightarrow\infty$, the $\omega^2$ terms vanish and $Z_\omega$ reduces to a Debye-like form, with its real and imaginary parts both being always negative.

In the low-frequency regime, where $\omega \ll \sqrt{ \omega_E \omega_{\rm pin} }$ and $\alpha \omega \ll \omega_{\rm pin}$ are simultaneously satisfied, $ X_\omega $ becomes approximately real and frequency-independent as
\begin{equation}
X_\omega = \frac{ 1 }{ \omega_{\rm pin} } \beta u_\omega .
\end{equation}
The DW thus oscillates in perfect synchronization with the ac current.
The emergent impedance $Z_\omega$ reduces to
\begin{equation}
Z_\omega = i \omega \left(\frac{\hbar\mu}{e}\right)^2 \frac{\gamma}{2\mu_0 M_{\rm S}\lambda S}
\left( - \frac{ \beta^2 }{ \omega_{\rm pin} } + \frac{ 1 }{ \omega_E } \right) ,
\label{z2}
\end{equation}
which is pure imaginary and scales linearly with $\omega$.
The phase of the induced SMF is therefore shifted by either $+\frac{\pi}{2}$ or $-\frac{\pi}{2}$ with respect to the applied ac current.
The sign of the phase shift is determined by the two competing contributions in the last bracket in Eq.~(\ref{z2}):
the $-\beta^2/\omega_{\rm pin}$ term, coming from the first term in Eq.~(\ref{smf}) and associated with the translational DW motion, and the $1/\omega_E$ term, resulting from the third term in Eq.~(\ref{smf}) and reflecting the time-dependent canting.
The former can be regarded as the fully dissipative contribution, i.e., the combined effect of the dissipative STT and dissipative emergent electric field, as reflected in the factor $\beta^2$, whereas the latter corresponds to the fully non-dissipative contribution.
At general frequencies, the non-dissipative and dissipative processes are mixed through their cross terms in Eq.~(\ref{z}).

\begin{figure}[t]
\centering
\includegraphics[width=8.5cm, bb=0 0 1623 649]{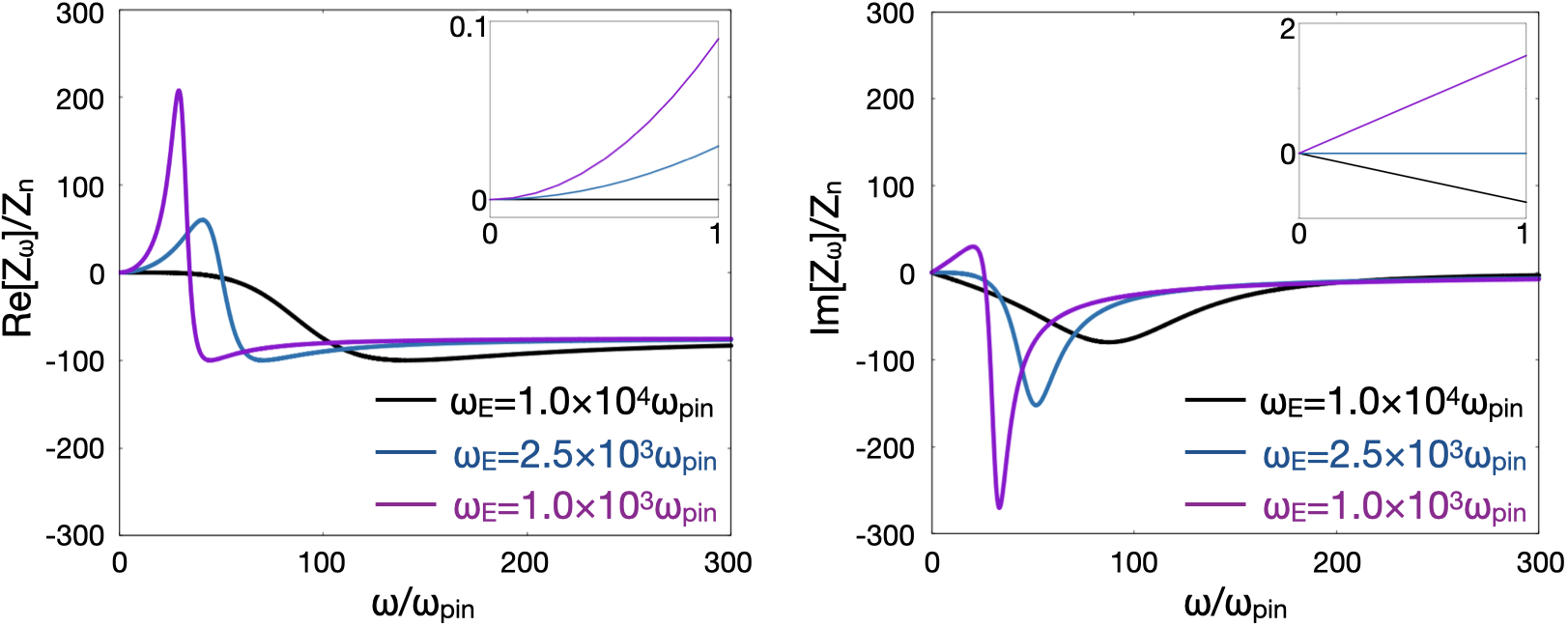}
\caption{
(a) Real and (b) imaginary parts of the emergent impedance $Z_\omega$, each normalized by $Z_n =\left(\frac{\hbar\mu}{e}\right)^2\frac{\gamma}{2\mu_0 M_S\lambda S}$.
The insets are magnified views of the frequency range of $0\leq\omega\leq1$.
We set $\alpha=0.01$ and $\beta=0.02$.
}
\end{figure}

In Fig.~2, the real and imaginary parts of $Z_\omega$ are plotted from Eq.~(\ref{z}) for three different values of $\omega_E$.
For the parameter values used, see the caption of the figure.
It is shown in the insets that, in the low-frequency limit, the real part ${\rm Re}[Z_\omega]$ vanishes, while the imaginary part ${\rm Im}[Z_\omega]$ scales linearly with $\omega$, consistent with Eq.~(\ref{z2}).
The slope of ${\rm Im}[Z_\omega]$ goes from positive to flat, and then to negative, as $\omega_E$ increases from $1.0\times10^3\omega_{\rm pin}$ to $2.5\times10^3\omega_{\rm pin}$, and then to $1.0\times10^4\omega_{\rm pin}$.
This is, again, understood from the competition between the $-\beta^2/\omega_{\rm pin}$ and $1/\omega_E$ terms in Eq.~(\ref{z2}):
$1/\omega_E-\beta^2/\omega_{\rm pin}>0$ ($<0$) for $\omega_E=1.0\times10^3\omega_{\rm pin}$ ($1.0\times10^4\omega_{\rm pin}$), leading to the positive (negative) slope of ${\rm Im}[Z_\omega]$.
For the case of $\omega_E=2.5\times10^3\omega_{\rm pin}$, $Z_\omega\simeq0$ in the low-frequency regime since $1/\omega_E-\beta^2/\omega_{\rm pin}=0$.

Such a qualitative $\omega_E$-dependence of the emergent impedance, where the sign of its imaginary part changes with varying $\omega_E$, offers a means to experimentally test the theoretical predictions.
For most AFM materials, however, $\omega_E$ is so dominant that $\beta^2/\omega_{\rm pin}\gg1/\omega_E$, and it is difficult to tune the value of $\omega_E$.
Typically, bulk AFMs exhibit $\frac{\omega_E}{2\pi}\sim10^{12}$ Hz \cite{textbook}.
The pinning potential $\omega_{\rm pin}$, on the other hand, is highly material-dependent and sensitive to disorder, temperature and other factors.
A rough estimate for $\omega_{\rm pin}$ can be obtained from the typical threshold magnetic field $B_{\rm th}$ for translational DW motion in Permalloy nanowires, a few tenths of millitesla \cite{hayashi}, which translates to $\frac{\omega_{\rm pin}}{2\pi}\sim \frac{\gamma B_{\rm th}}{2\pi}\sim10^6$ Hz with $\gamma=1.76\times10^{11}$ Hz/T.
Thus, the canting-induced contribution is relevant if $\beta\sim0.001$ or smaller. 
The recent experimental observation of emergent impedance in a bulk AFM FeSn$_2$, which focused on the low-frequency regime ($\sim$kHz), can indeed be explained by the translational DW motion \cite{yokouchi-arxiv}.
Here, we propose that synthetic AFMs provide particularly promising platforms for demonstrating the predicted $\omega_E$-dependence of the emergent impedance.
In these systems, $\omega_E$ corresponds to the interlayer exchange coupling between two ferromagnetic layers, and can be tuned by adjusting the thickness or material properties of the nonmagnetic spacer layer \cite{grunberg}.
Achieving reasonably small values of $\omega_E$ in synthetic AFMs will allow probing the emergent impedance arising from the time-dependent canting moment.

\section{Conclusion}
In conclusion, we have theoretically investigated the emergent impedance induced by antiferromagnetic domain-wall dynamics.
While the contribution from the translational domain-wall motion resembles its ferromagnetic counterpart, the time-dependent canting moment produces effects unique to antiferromagnets.
In particular, the emergent impedance arising from the time-dependent canting moment is inversely proportional to the antiferromagnetic exchange coupling constant, and competes with that generated by the translational motion.
These findings reveal the fundamental characteristics of emergent impedance in antiferromagnets, providing a foundation for future studies of their electrical response and potential spintronics applications.

\section{Acknowledgments}
The authors thank T. Yokouchi and Y. Shiomi for fruitful discussions.
This work was supported by Japan Society for the Promotion of Science KAKENHI (nos. 22K03538, 23K17882, 23K26521, 24H00039, and 24H00409), and Japan Science and Technology Agency TI-FRIS program.

\section{Data Availability}
The data that support the findings of this article are not publicly available.
The data are available from the authors upon reasonable request.


\end{document}